\documentclass[pre,aps,twocolumn]{revtex4}

\usepackage{graphicx}
\usepackage{amsmath,empheq}
\usepackage{amssymb}
\usepackage{ifthen}
\usepackage{booktabs}

\usepackage{graphicx}% Include figure files
\usepackage{dcolumn}% Align table columns on decimal point
\usepackage{bm}% bold math
\usepackage{longtable}
\usepackage{gensymb}

\newcommand{\bb}{\begin{equation}}
\newcommand{\ee}{\end{equation}}
\newcommand{\ba}{\begin{eqnarray*}}
\newcommand{\ea}{\end{eqnarray*}}
\newcommand{\rhor}{\rho({\bf r})}

\newcommand{\rr}{{\mathbf r}}

\begin{document}

\title{Modified Kelvin equations for capillary condensation in narrow and wide grooves}

\author{Alexandr \surname{Malijevsk\'y}}
\affiliation{
{Department of Physical Chemistry, University of Chemical Technology Prague, Praha 6, 166 28, Czech Republic;}\\
 {Department of Molecular and Mesoscopic Modelling, ICPF of the Czech Academy Sciences, Prague, Czech Republic}}                %Institute of Chemical Process Fundamentals of the CAS, v. v. i., Prague, Czech Republic}}
\author{Andrew O. \surname{Parry}}
\affiliation{Department of Mathematics, Imperial College London, London SW7 2BZ, UK}

\begin{abstract}
\noindent We consider the location and order of capillary condensation transitions occurring in deep grooves of width $L$ and depth $D$. For walls
that are completely wet by liquid (contact angle $\theta=0$) the transition is continuous and its location is not sensitive to the depth of the
groove. However for walls which are partially wet by liquid, where the transition is first-order, we show that the pressure at which it occurs is
determined by a modified Kelvin equation characterized by an edge contact angle $\theta_E$ describing the shape of the meniscus formed at the top of
the groove. The dependence of $\theta_E$ on the groove depth $D$ relies, in turn, on whether corner menisci are formed at the bottom of the groove in
the low density gas-like phase. While for macroscopically wide grooves these are always present when $\theta<45\degree$ we argue that their formation
is inhibited in narrow grooves. This has a number of implications including that the local pining of the meniscus and location of the condensation
transition is different depending on whether the contact angle is greater or less than a universal value $\theta^*\approx 31\degree$. Our arguments
are supported by detailed microscopic density functional theory calculations which show that the modified Kelvin equation remains highly accurate
even when $L$ and $D$ are of the order of tens of molecular diameters.
\end{abstract}

\maketitle

\begin{figure}
\includegraphics[width=3.5cm]{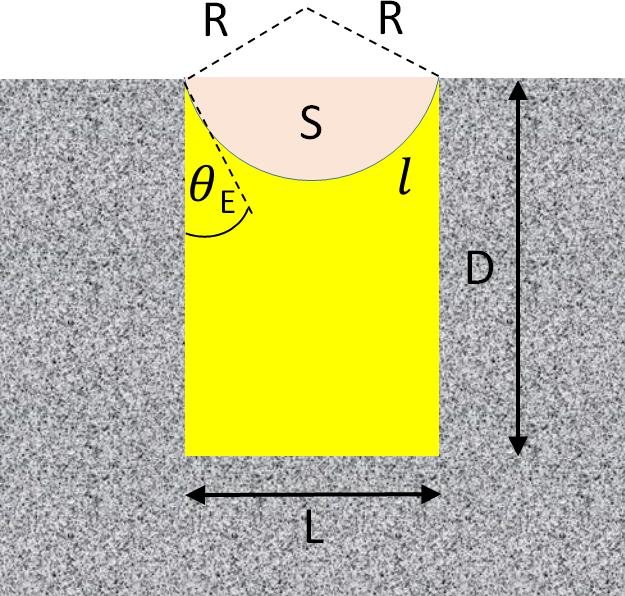} \hspace*{0.5cm} \includegraphics[width=3.5cm]{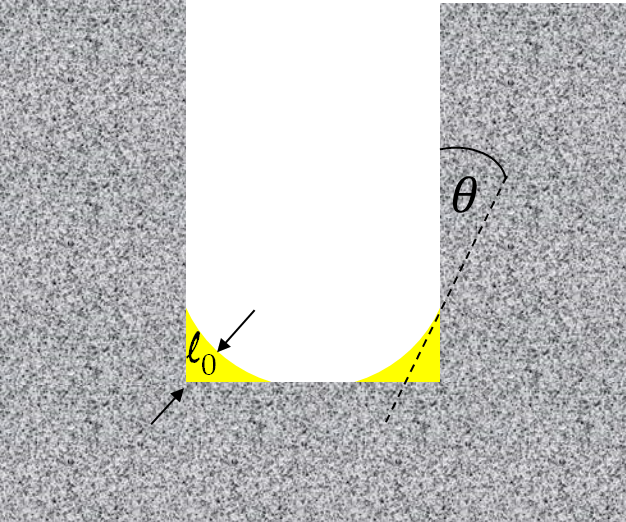}
\caption{Left: Schematic cross-section of a wide groove showing the meniscus of radius $R$ pinned to near the top, with edge contact angle
$\theta_E$, characterizing the condensed liquid-like (yellow) phase. Symbols are described in the text. Right: In the gas-like phase, corner menisci
of liquid (yellow) of thickness $\ell_0$ nucleate near the groove bottom and meet the side and bottom walls at the equilibrium contact angle
$\theta$, when the contact angle $\theta<\pi/4$ and the width is sufficiently large.}
\end{figure}

The equilibrium and dynamical properties of fluids are strongly affected by geometric confinement inducing new first-order and continuous phase
transitions such as wetting, pre-wetting and wedge (corner) filling \cite{dietrich, schick, hend, rejmer, wood1, binder03, bonn, our_prl, our_wedge}.
These transitions together with the ubiquitous and much studied phenomenon of capillary condensation can combine to produce very rich phase diagrams
even for fluids confined by relatively simple geometrical structures. In this paper we revisit the condensation of a fluid occurring in a capped
capillary slit or equivalently in a macroscopically long groove of depth $D$ and width $L$ scored in a solid surface in contact with a bulk vapour.
This has attracted considerable interest in the last decade where it has been shown, for example, that the condensation, that is the filling of the
groove as the pressure is increased, is continuous when the walls are completely wet by liquid \cite{evans_cc, tasin, mistura, hofmann, mal_groove,
parry_groove, ser, mal_13, our_groove, our_groove_lett}. Here we revisit the first-order condensation occurring when the walls are partially wet
corresponding to contact angle $\theta>0$. Similar to recent studies of adsorption in fully open pores \cite{fine_slit} we first show, using
macroscopic arguments, that the location of the condensation in a groove of {\it{finite depth}} is determined by a modified Kelvin equation which
depends on the value of an edge contact angle $\theta_E$ describing the shape of the meniscus at the top. The dependence of $\theta_E$ on the contact
angle $\theta$ and groove aspect ratio $L/D$ is determined and shown to fall into two two regimes, depending on whether corner menisci are formed in
the gas-like phase at condensation. Whilst such corner menisci are always present in macroscopically wide grooves when $\theta<\pi/4$, we argue that
their formation is suppressed in microscopically narrow slits even for small contact angles. A direct consequence of this is that for nanoscopically
narrow grooves there should be a strong qualitative change in the location of the condensation when the contact angle approaches a universal value
close to $31\degree$ at which the local pinning of the meniscus at the top vanishes. We have tested this using a fully microscopic density functional
theory (DFT) and the results are in excellent agreement with the theoretical predictions over a wide range of aspect ratios $L/D$.

To begin we use macroscopic arguments to derive a modified Kelvin equation determining the pressure shift $\delta p_{cc}(L;D)=p_{sat}-p_{cc}(L;D)$,
relative to bulk saturation, at which a gas-like phase condenses to a liquid-like phase in a macroscopically long, deep capillary groove of width $L$
and depth $D\gg L$ which we consider in contact with a bulk reservoir of vapour. Recall that for an infinitely deep groove $\delta p_{cc}(L;\infty)=
\delta p_{\rm Kel}(L)$ where $\delta p_{\rm Kel}(L)=2\gamma\cos\theta/L$ is the standard Kelvin equation prediction for the shift from bulk
coexistence and $\gamma$ is the liquid-gas surface tension \cite{tar_ev, evans90}. This macroscopic prediction, which is based on simply balancing
the Grand Potentials of coexisting liquid-like and gas-like phases, is remarkably robust and for partial wetting in particular, accurately determines
the location of the condensation down to slit widths on the nanoscopic scale \cite{tar_ev, evans90}. Let us now generalize the derivation of the
Kelvin equation to grooves of finite depth $D$. To this end consider the grand potential $\Omega_l$, per unit length of the groove, of a dense
liquid-like phase at the same chemical potential as the gas reservoir. This state, which is metastable in bulk, has pressure $p^\dagger=p-\delta p$
with $\delta p>0$. In this case the groove is filled with liquid and a meniscus separating the liquid from the gas reservoir, at pressure $p$, must
be pinned at the top of the groove (see Fig.~1). Macroscopically, the meniscus must be of circular cross-section, with Laplace radius
$R=\gamma/\delta p$, and meet the top side walls of the groove at an edge contact angle $\theta_E$. Notice that in general this will be
{\it{different}} from Young's contact angle $\theta$ since the pinning of the meniscus at the top no longer necessitates the balance of surface
tension forces which determines the value of $\theta$ at a planar wall. The edge contact angle and Young's contact angle are only the same if the
pressure at condensation happens to be identical to that occurring in an infinitely deep slit since it is only in that case that a circular meniscus
can be accommodated in the groove that meets the walls at Young's contact angle; if $\delta p_{cc}(L;D)\ne \delta p_{\rm Kel}(L)$ this cannot happen
and the meniscus is forced to the top of the groove where it is free to have a different angle of contact. Thus for the liquid-like phase we can
write
\begin{equation}
\Omega_l\approx-p^\dagger(LD-S)-pS+\gamma_{wl}(2D+L)+\gamma\ell
\end{equation}
Here $\gamma_{wl}$ is the wall-liquid surface tension while $S=(\pi/2-\theta_E)R^2-\sin\theta_ERL/2$ is the area between the meniscus and the open
end and $\ell=(\pi-2\theta_E)R$ is the meniscus length (see Fig.~1). We have assumed that the groove is sufficiently deep, $D>L/2$, so that the
meniscus does not touch the bottom wall for any value of the contact angle. A similar argument determines the grand potential $\Omega_g$, per unit
length for the gas-like phase. However, here we need to distinguish between two scenarios since the two corners at the bottom of the groove may
themselves locally nucleate a liquid-like phase forming corner menisci. Macroscopically this can only happen for contact angles
$\theta<\frac{\pi}{4}$ which corresponds to the phase boundary for the wedge filling transition at a right angle corner \cite{fill1, fill2, fill3}.
For $\theta>\frac{\pi}{4}$ no corner menisci are formed regardless of the width of the slit and the value of the pressure. In this case we can write
$\Omega_g\approx -pLD+\gamma_{wg}(2D+L)$ where $\gamma_{wg}$ is the wall-gas surface tension. However, for $\theta<\frac{\pi}{4}$ this is modified to
\begin{equation}
\Omega_g\approx -pLD+\gamma_{wg}(2D+L)+2\Delta\Omega_{cm} \label{Omegacm}
\end{equation}
where $\Delta\Omega_{cm}$ is the contribution coming from the two circular corner menisci. The size and location of these are determined
geometrically from noting that they are of radius $R=\gamma/\delta p$ and must touch both the side and bottom walls at the equilibrium contact angle
$\theta$. This determines that the local thickness of each corner meniscus, as measured from the appropriate corner, is
$\ell_0=R(\sqrt{2}\cos\theta-1)$ and that the increment to the free-energy is given by
 \bb \Delta\Omega_{cm}=-\gamma
 R\left[\sqrt{2}\cos\theta\sin\left(\frac{\pi}{4}-\theta\right)-\frac{\pi}{4}+\theta\right]\,,
 \ee
which simply arises from considering the area and length of each meniscus. Note that $\ell_0$ and $\Omega_{cm}$ vanish as $\theta$ is increased to
$\pi/4$ corresponding to the filling phase boundary for the right angle corner \cite{fill1,fill2,fill3}. The corner menisci remain separate unless
$\theta=0$ in which case they merge precisely at the pressure of condensation \cite{parry_groove}.

Capillary condensation occurs when the grand potentials of each phase balance, $\Omega_l=\Omega_g$, which determines that the value of the pressure
is described by the modified Kelvin equation
\begin{equation}
\delta p_{cc}(L;D)=\frac{2\gamma\cos\theta_E}{L}\,, \label{kelvin}
\end{equation}
which is just the geometrical condition that the meniscus spans the capillary groove. The value of the edge contact angle, or rather its dependence
on $\theta$ and the aspect ratio $L/D$, is determined self consistently, from solving
 \bb
\cos\theta_E-\cos\theta=\alpha(\theta,\theta_E)\frac{L}{4D}\,. \label{thetaE}
 \ee
There are two regimes: For $\theta>\frac{\pi}{4}$ (corner menisci absent)
 \bb
 \alpha(\theta,\theta_E)=\frac{\pi/2-\theta_E}{\cos\theta_E}+\sin\theta_E-2\cos\theta\,,\label{ct1}
 \ee
 while for $\theta<\frac{\pi}{4}$ (corner menisci present)
 \bb
 \alpha(\theta,\theta_E)=\frac{\sin\theta_E}{2}-\cos\theta+\frac{\cos\theta(\cos\theta-\sin\theta)+\theta-\frac{\theta_E}{2}}{\cos\theta_E}\,.
\label{ct2}
 \ee

For complete wetting the only solution of these equations, for any aspect ratio $L/D$, is $\theta_E=\theta=0$. In other words for walls that are
completely wet the Kelvin equation is not altered by the groove depth and $\delta p_{cc}(L;D)=\delta p_{\rm Kel}=2\gamma/L$. Notice that the
condition $\theta_E=\theta$ means there is no local pinning of the meniscus at the top of the groove since, at condensation, it may be translated up
and down the groove without any cost of energy; at the pressure of condensation a circular meniscus, touching the side walls tangentially, can exist
at any height in the groove including the very top. Similarly there is no pinning at the bottom since the two corner menisci are both of radius $L/2$
meaning they have already merged to form a single meniscus -- no other interfacial configuration characterizing the gas-like phase has a lower free
energy. The absence of pinning at either end are necessary conditions that, for the case of complete wetting, the condensation transition is
continuous for all depths $D$. This is not altered by including intermolecular forces which, again for the case of complete wetting, repel the
meniscus from both the groove top and bottom \cite{evans_cc}.

We now turn to partial wetting. In the limit $D\to\infty$, corresponding to an infinitely deep groove, $\theta_E\to\theta$ and consequently $\delta
p_{cc}(L;D)\to\delta p_{\rm Kel}(L)=2\gamma\cos\theta/L$ as required. For all finite $D$, however, the values of the edge and Young's contact angle
are different, with $\theta_E>\theta$, and the meniscus is pinned near the top of the groove. It is possible to consider constrained configurations
where the meniscus is delocalized, away from the groove edge, but these have a higher free energy. Similarly at the pressure of condensation the
corner menisci, characterizing the gas-like phase, are pinned to each bottom since their radius $R<L/2$, and it costs free energy to merge them into
a single meniscus spanning the groove. Such local pinning means that we can be certain that for partial wetting the condensation transition in a deep
groove is first-order.

The above macroscopic analysis implies that for all finite depth capillaries the edge contact angle is greater than Young's contact angle; the
pressure at which capillary condensation occurs is closer to saturation than for the infinite slit i.e $\delta p_{cc}(L,D)<\delta p_{\rm Kel}(L)$.
Let us quantify this by considering the leading-order correction to the prediction of the standard Kelvin equation occurring for deep grooves by
writing
 \bb
 \cos\theta_E=\cos\theta-\frac{L}{4D}\tilde{\alpha}(\theta)+\cdots \label{asymp}
 \ee
 We immediately identify the coefficient $\tilde{\alpha}(\theta)=-\alpha(\theta,\theta)$ the value for which again falls into two regimes. For
$\theta>\pi/4$ we have
 \bb
 \tilde{\alpha}(\theta)=\left(\frac{\pi}{2}-\theta\right)\sec\theta+\sin\theta-2\cos\theta \label{alpha*}
  \ee
while for $\theta<\pi/4$ the presence of the corner menisci modifies this to $\tilde{\alpha}(\theta)=\theta\sec\theta-\sin\theta$. The function
$\tilde{\alpha}(\theta)$ has limiting values $\tilde{\alpha}(0)=0$, since there is no modification to the standard Kelvin equation for complete
wetting, and $\tilde{\alpha}(\pi/2)=2$. In each regime $\tilde{\alpha}(\theta)$ decreases monotonically as $\theta$ decreases and is continuous and
differentiable at $\theta=\pi/4$, exhibiting a jump in its second derivative at this point. The analysis gives a consistent account of both the
pressure and order of the condensation in deep wide grooves; for complete wetting the transition is continuous occurring  at a value of the pressure
$p=p_{\rm Kel}(L)$, which is independent of the groove depth, while for partial wetting the transition is first-order occurring at a shifted value
$p=p_{cc}(L;D)$ which is closer to saturation than $p=p_{\rm Kel}(L)$ since $\theta_E>\theta$. This can be rationalized since, similar to a slit
which is open at both ends, it costs free energy to form the meniscus in the liquid-like phase.

However, this macroscopic analysis also points to an intriguing possibility. Imagine that the formation of the corner menisci was somehow suppressed
so that Eq.~(\ref{alpha*}) determines $\tilde{\alpha}(\theta)$ over the whole range of contact angles including the low contact angle region
$\theta<\pi/4$. If this were the case then $\tilde{\alpha}(\theta)$ falls into two regimes on either side of a new contact angle $\theta^*$
determined by $\tilde{\alpha}(\theta^*)=0$; for $\theta>\theta^*$, where $\theta^*\approx 31\degree$, the edge contact angle $\theta_E>\theta$ while
for $\theta<\theta^*$, $\theta_E<\theta$. The value of $\theta^*$ is universal. Hence $\theta_E=\theta\approx 31\degree$ for all aspect ratios $L/D$.
A direct consequence is that $\delta p_{cc}(L;D)>\delta p_{\rm Kel}(L)$ so that, counter intuitively, condensation occurs further from saturation
than for the infinite slit despite the fact that a meniscus must be created in the liquid-like phase.

One simple mechanism that can lead to this scenario is to consider condensation occurring in narrow grooves of order tens of molecular diameters.
Recall that for partial wetting the macroscopic analysis, leading to the Kelvin equation for infinitely deep groove, predicts accurately the location
of the condensation transition for slits which are of this width \cite{evans90}. We can therefore reasonably expect that the above generalized
equations work well for such narrow systems when the depth is finite. However for such narrow slits, the shift of the condensation transition away
from bulk coexistence may be so large that corner menisci cannot form in the gas-phase even if $\theta<\pi/4$. That is, their characteristic size,
$\ell_{0}$, as predicted by the above mesoscopic analysis is not substantially larger than the underlying molecular diameter $\sigma$. For example,
using the result for $\ell_0$ quoted above means that for $\theta\approx 30\degree$ we need to have $L\gg 8\,\sigma$ in order that $\ell_0\gg\sigma$.
If this condition is not met it is reasonable to conjecture that the macroscopic analysis should be based on the assumption that there is no
contribution $\Delta\Omega_{cm}$ in the grand potential (\ref{Omegacm}) of the gas-like phase. The analysis of the free-energy of the liquid-like
phase is not affected since an interface or meniscus separating the capillary liquid from the outer bulk gas must always be present regardless of the
slit width $L$ (below the temperature corresponding to the capillary critical point). It is natural therefore to hypothesize that when $L$ is only
tens of atomic diameters and $\theta<\theta^*$  condensation in a finite depth groove occurs further away from saturation compared to the prediction
$\delta p_{\rm Kel}(L)$.

In order to test this hypothesis we turn to a microscopic DFT \cite{evans79} based on minimization of a model grand potential functional of the
averaged one-body density of the fluid $\rhor$:
\begin{equation}
\Omega[\rho]=F[\rho]-\int d{\bf{r}}(\mu-V({\bf{r}}))\rho({\bf{r}}) \label{omega}
\end{equation}
where $V({\bf{r}})$ is the external potential due to the confining walls of the groove. The model intrinsic Helmholtz free-energy functional
$F[\rho]$ contains all the information about the fluid - see the supplementary material for further details. For this we use a highly accurate
Rosenfeld hard-sphere functional \cite{ros,mal_13} to modelling packing effects associated with volume exclusion, and a reliable mean-field treatment
of the attractive interaction $u_{\rm att}(r)$ between the fluid atoms.  The external potential $V(\rr)=V(x,z)$  arises from the presence of the
groove walls placed in the plane $z=0$ (bottom), $z=D$ (top) and  $x=-L/2$ and $x=L/2$ (sides). All the walls are assumed to be infinitely long, so
that the system is translation invariant along the $y$ axis. The walls are formed from a uniform distribution of atoms with density $\rho_w$, each
interacting with the fluid atoms via a Lennard-Jones $12$-$6$ potential $\phi_w(r)$, so the net external potential is obtained by integrating
$\phi_w$ over the whole volume domain of the walls (see Supplementary materials for details). For this realistic model fluid the present mean-field
DFT should determine accurately the location of the first-order condensation transition. We note that standard finite-size scaling considerations
suggest that beyond mean-field the transition is very sharply rounded on a scale of order $\exp[-\gamma LD/k_BT]$ but, away from the immediate
vicinity of the (capillary) critical point, this is completely negligible.

\begin{figure}
\includegraphics[width=4cm]{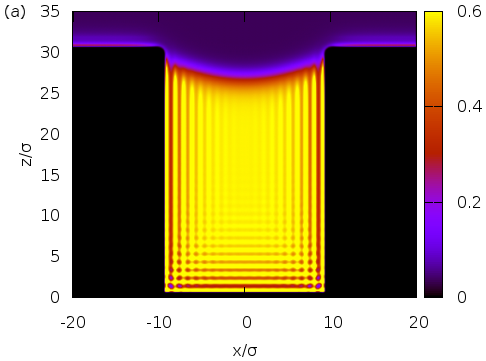} \hspace*{0.2cm} \includegraphics[width=4cm]{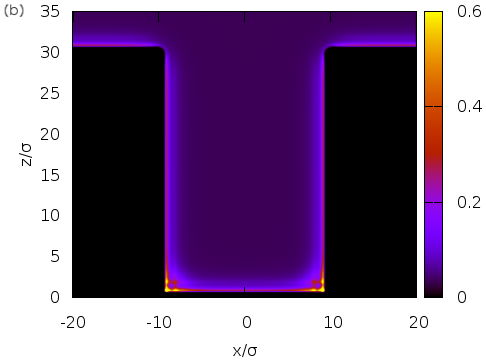}
\caption{Equilibrium density profiles at the condensation transition of coexisting (a) liquid-like  and (b) gas-like phases in the groove of width
$L=20\,\sigma$ and depth $D=30\,\sigma$ at a temperature $T=0.71\,T_c$.}
\end{figure}

Here we consider groove walls in the partial wetting regime, so that the temperature of the system is always below that of the (first-order) wetting
transition occurring at $T_w=0.8\,T_c$, where $T_c$ is the bulk critical temperature of the fluid. We vary the depth of the grooves fixing the width
to $L=20\,\sigma$ which we expect to be sufficiently small to prevent the formation of the corner menisci. Indeed, according to our estimation of
$\ell_0$ above, the thickness of the adsorbed menisci would be $\ell_0\approx 2\,\sigma$ which is clearly indistinguishable from the local adsorption
of an atom at the corner. The absence of corner menisci for such narrow grooves is confirmed by our DFT results, see Fig.~2, showing the coexisting
states at capillary condensation for $\theta\approx35\degree$ in a groove of depth $D=30\,\sigma$. Although the upper meniscus characterizing the
high-density state is well pronounced (Fig.~2a), only a very small microscopic enhancement of the adsorption, akin to simple local corner layering,
is present at the bottom in the low-density, gas-like state (Fig.~2b).

\begin{figure}
\centerline{\includegraphics[width=4.4cm]{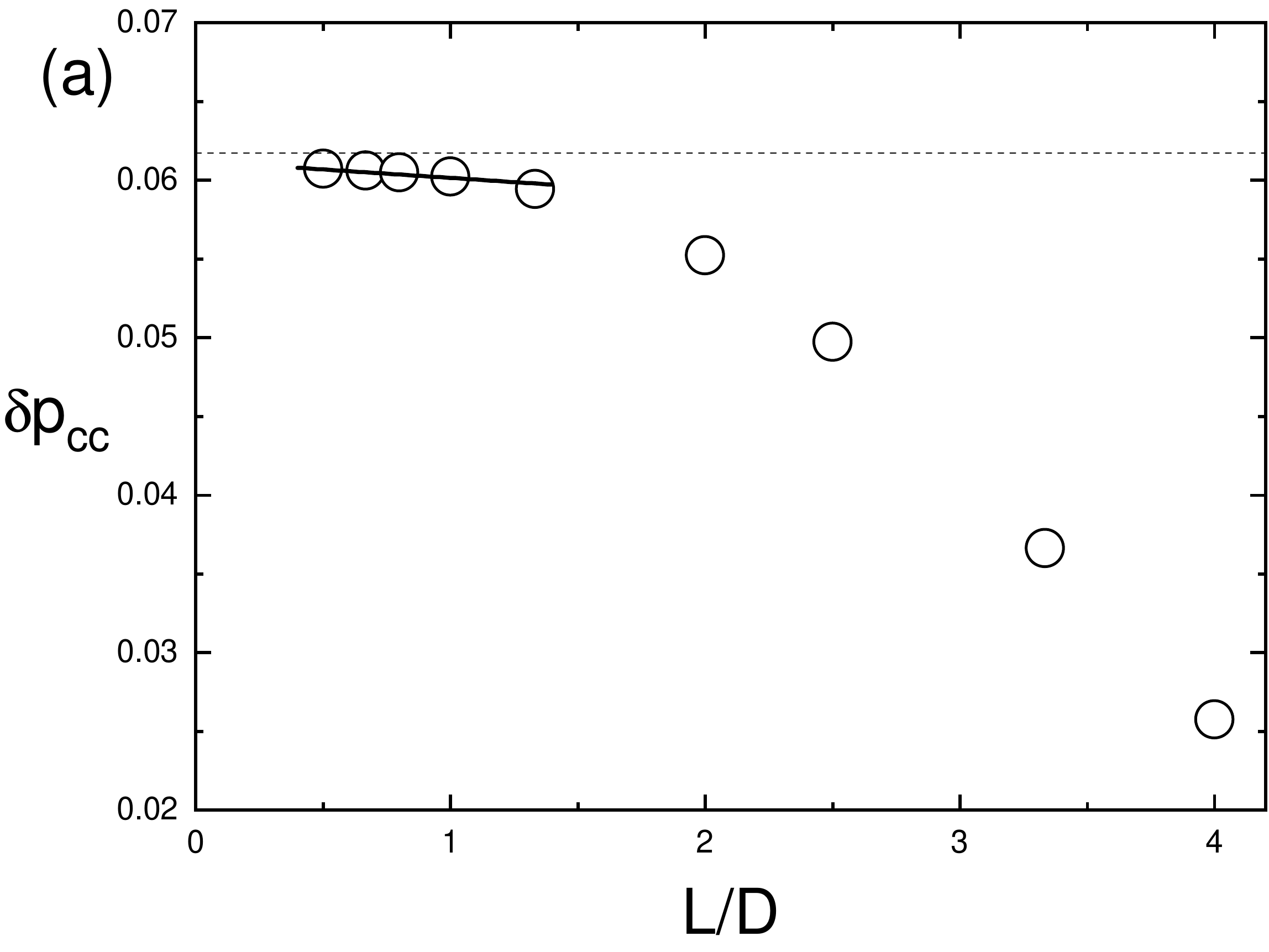} \hspace*{0.1cm} \includegraphics[width=4.4cm]{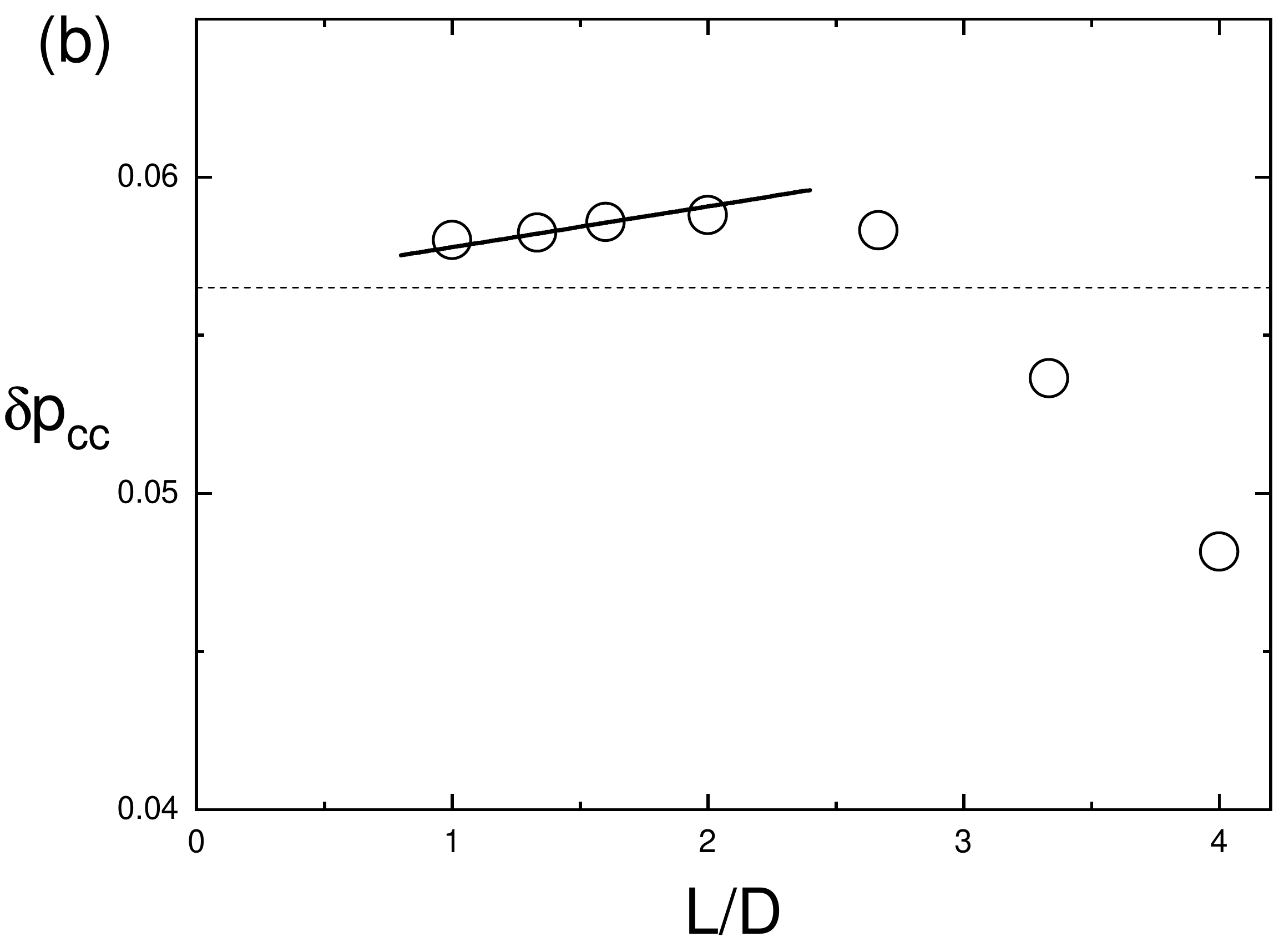}}
 \caption{DFT results for the shift $\delta p_{cc}(L;D)=p_{\rm sat}-p_{cc}(L;D)$ (in units of $\varepsilon/\sigma^3$)
in location of the  condensation transition in a groove of width $L=20\,\sigma$ as a function of the aspect ratio $L/D$ at two different
temperatures: (a) $T=0.71\,T_c$ corresponding to $\theta\approx 35\degree$
 and (b) $T=0.78\,T_c$ corresponding to $\theta\approx 20\degree$. In each case the solid curve is the asymptotic result obtained analytically
from Eqs.~(\ref{kelvin}), (\ref{asymp}) and (\ref{alpha*}), inputting the contact angles $\theta$ obtained from DFT. In each figure the dashed line
refers to the pressure of condensation in the infinitely deep slit $L/D=0$ which we have determined independently.}\label{dft_res}
\end{figure}

These density profiles suggest that for such narrow, deep grooves the modification of $\theta_E$ from $\theta$ (and thus the shift $\delta
p_{cc}(L;D)$) should be governed by the function $\tilde{\alpha}(\theta)$ given in Eq.~(\ref{alpha*}), although $\theta<45\degree$. We have tested
this by determining $\delta p_{cc}(L;D)$  for grooves of different depths, and at two different temperatures corresponding to contact angles above
and below the predicted value $\theta^*\approx 31\degree$ (see Fig.~\ref{dft_res}). In both cases as $D$ increases the value of $\delta p_{cc}(L;D)$
approaches a constant which lies very close to the value $\delta p_{\rm Kel}$ predicted by the standard Kelvin equation. However for the lower
temperature $T=0.71\,T_c$ (corresponding to $\theta\approx 35\degree$) we observe that $\delta p_{cc}(L;D)$ approaches the limiting value from below,
while for $T=0.78\,T_c$ (corresponding to $\theta\approx 20 \degree$) $\delta p_{cc}(L;D)$ approaches the limiting value from above in keeping with
the predicted change in sign of $\tilde{\alpha}(\theta)$ near $\theta^*\approx 31\degree$. As is apparent, the first-order linearization of the
modified Kelvin equation (\ref{asymp}) together with Eq.~(\ref{alpha*}) very accurately determines the pressure of the condensation transition even
for aspect ratios $L/D$ of order unity. Indeed the results for $\delta p_{cc}(L;D)$  deviate from the linear result (\ref{asymp}) only when
$L/D\approx 2$ which is precisely when we anticipate that the meniscus starts interacting with the groove bottom.

The possibility that the edge contact angle $\theta_E$ can be less than Young's contact angle and describes accurately the pressure of the
condensation transition in the (modified) Kelvin equation for narrow slits is the main result of our paper. There are further consequences. For
example if instead of fixing $T$ and varying $D$ we fix the depth $D$ and increase the temperature, Eqs.~(\ref{thetaE}) and (\ref{ct2}) predict that
$\theta_E$ vanishes, for a deep groove, when $\theta\approx 0.78 \sqrt{L/D}$ -- that is the edge contact angle vanishes even though the walls remain
partially wet by liquid. This corresponds to an apparent, first-order, wetting transition occurring below the true wetting temperature $T_w$ for an
infinite area wall (at which $\theta$ vanishes). Our results generalize in a number of ways. For example, we can also consider a slightly modified
groove where the side walls have contact angle $\theta$ and the bottom wall has a different contact angle $\theta_B$ \cite{parry_groove}. In this
case the final cosine on the R.H.S. of Eq.~(\ref{alpha*}) is replaced with $\cos\theta_B$. Consequently the value of $\theta^*$ at which
$\theta_E=\theta=\theta^*$, and the local pinning at the top vanishes, is altered. This establishes that when the side walls are neutral
($\theta=\pi/2$), the pinning is absent when $\theta_B=0$. In other words the pinning vanishes precisely at the wetting temperature of the wall at
the groove bottom.

In summary, we have shown the the pressure of the condensation transition in a capillary groove is determined by a modified Kelvin equation
characterized by an edge contact angle $\theta_E$ which, for partial wetting, depends on the geometrical aspect ratio $L/D$. For narrow slits the
suppression of corner menisci in the gas-like phase at condensation implies that  $\theta_E$ is less than Young's contact angle provided $\theta$ is
less than a universal value $\theta^*\approx 31\degree$. This leads to a qualitative change in the pressure of the condensation transition which is
confirmed by a detailed microscopic DFT study. Similar phenomena also occur in other geometries such as a cylindrical pore of finite depth.

%\section{Summary}

\begin{acknowledgments}
\noindent This work was funded in part by the EPSRC UK grant EP/L020564/1, ``Multiscale Analysis of Complex Interfacial Phenomena''.
 A.M. acknowledges the support from the Czech Science Foundation, project 17-25100S.
\end{acknowledgments}

\end{document}